\documentclass[copyright,creativecommons]{eptcs}
\providecommand{\event}{}
\providecommand{\volume}{2}
\providecommand{\anno}{2009}
\providecommand{\firstpage}{67}
\usepackage{breakurl}

\usepackage{pstricks,pst-node,pst-tree}
\usepackage{graphicx}
\usepackage{color}
\usepackage{latexsym}
\usepackage{stmaryrd}

\title{Reasoning About a Service-oriented Programming Paradigm}
\author{Claudio Guidi$^{1}$
\institute{Department of Computer Science, University of Bologna, Italy}
\email{cguidi@cs.unibo.it}
\and
Fabrizio Montesi$^{1}$
\institute{italianaSoftware s.r.l., Italy}
\email{fmontesi@italianasoftware.com}
}
\footnotetext[1]{These authors contributed equally to this work.}
\def\titlerunning{Reasoning About a Service-oriented Programming Paradigm}
\def\authorrunning{C.~Guidi \& F.~Montesi}
\begin{document}
\maketitle

\begin{abstract}
This paper is about a new way for programming distributed applications: the service-oriented one. It is a concept paper based upon our experience in developing a theory and a language for programming services. Both the theoretical formalization and the language interpreter showed us the evidence that a new programming paradigm exists. In this paper we illustrate the basic features it is characterized by.
\end{abstract}

\section{Introduction}
This paper is about a new way for programming distributed applications: the
service-oriented one.
It is a concept paper based upon our experience in developing a theory and a
language for programming services. Our work started some years ago when we began
to formalize the basic mechanisms of the Web Services technology in a process
calculus. We chose such an approach because process calculi were naturally born
for describing concurrent processes, as Web Services are.
The need to address Web Services with a formal approach was motivated by the
high level of complexity they are characterized by: we wanted a simple and
precise means for catching their essentials and, at the same time, strong
foundations for developing concrete tools for designing and implementing service systems.

When we started to build our formal model we took inspiration from foundational calculi such as CCS
\cite{ccs} and $\pi$-calculus \cite{pi}, by enriching those approaches with specific mechanisms which came from the Web Services technology.
We developed SOCK \cite{icsoc06,myphd}, that is a process calculus where
those aspects of service-oriented computing which deal with communication primitives, work-flow composition, service session management and service networks are considered. SOCK is structured on three
layers, each one representing a specific feature of the service-oriented
approach: the \emph{behaviour} of a service, the execution of service
\emph{sessions} into a \emph{service engine} and the connection of services within a \emph{network}.
Such a categorization allowed us to handle the complexity of service-oriented
computing without loosing the important details they are characterize by.
Differently from our approach, other authors proposed SCC \cite{scc} and COWS \cite{ws-calculus}. The main difference between SCC and SOCK can be found in the session identification mechanism. In SOCK we identify sessions by means of \emph{correlation sets} whereas in SCC sessions are identified by freshly generated names. In our opinion, correlation set represents a key mechanism for the service-oriented programming paradigm which is also modelled in COWS and it is provided by the most credited orchestration language for Web Services: WS-BPEL \cite{wsbpel-spec}. The main difference between SOCK and COWS is the fact that SOCK explicitly supports a state whereas COWS does not. Moroever, SOCK encoded the RequestResponse communication primitive which is not supported by COWS. Both state and the RequestResponse primitive made SOCK close to the technologies. This fact allowed us to reason about fault handling issues which led us to propose a new way, the dynamic handling, for managing faults \cite{acsd08,ecows08}.

\newpage
At the time we created SOCK, we did not imagine that a new way for approaching distributed
system design could be developed starting from it.
The evidence of this came some years later, when we started to develop the JOLIE
programming language \cite{JolieSite,CoOrg06}.
JOLIE was born as a strict implementation of the semantics contained in SOCK.
The syntax was inspired by that of the process calculus, blended with some
common constructs which are familiar to those accustomed to languages such as C
and Java. The JOLIE language allowed us to apply the concepts studied in SOCK in
real world application development, and this raised new issues and in turn made
new software design patterns to emerge. Addressing these issues brought to the
definition of new mechanisms for service system composition such as
\emph{embedding} and \emph{redirecting}. 
Embedding increases the granularity level of services into a system, whereas redirecting allows for grouping services under a unique endpoint. Both the approaches can be freely mixed together in order to obtain new systems of services.
Today, at the best of our knowledge, JOLIE is the first language which allows for the designing of a distributed system completely composed by services.



At the present, we can state that SOCK and JOLIE form a framework that offers
the possibility to study service-oriented computing issues from the
theoretical, architectural and practical points of view. The sum of these
experiences made us aware of the fact that we were facing a new way for
designing, developing and studying distributed applications: the
service-oriented paradigm. It is difficult and ambitious to highlight the
distinctive features of a new programming paradigm, but with this paper we would
like to share our strong impressions that the service-oriented paradigm exists.

In the following we describe the basic concepts that we have extracted and we
try to show how they can be considered the foundations of the service-oriented
programming paradigm. Our work is strongly influenced by that made by the
industrial and scientific communities on service-oriented computing, and strives
to consider the most relevant results in our definition of the service-oriented
approach. In section \ref{sec:service} we present our definition of \emph{service}, after introducing the main concepts that are behind it. In section \ref{sec:composition} we expose how one can compose services in order to obtain a distributed system in which they communicate with each other. Section \ref{sec:discussion} contains some design pattern examples and remarks that emerged from our experience in using the programming paradigm we propose. Finally, in section \ref{sec:conclusions} we report our conclusions and future works.


\section{Definition of Service}\label{sec:service}
The first thing to address is the definition of the term \emph{service}. In order to provide evidence that a service-oriented paradigm exists we need to define what a service is, as services are the most important component of the paradigm. The definition of service for the W3C Working Group~\cite{WebServiceGlossary} follows:
\begin{quote}
\emph{``A service is an abstract resource that represents a capability of performing tasks that form a coherent functionality from the point of view of providers entities and requesters entities. To be used, a service must be realized by a concrete provider agent.''}
\end{quote}
We agree with this definition but we argue that it is too abstract because too many things could be a service. At the end of this section we present our definition of service, which is based upon the concepts of \emph{behaviour}, \emph{engine} and service \emph{description}.

\subsection{Behaviour}\label{sec:behaviour}
Here we present the definition of \emph{behaviour} of a service, which introduces two basic concepts: service \emph{activities} and their \emph{composition} in a work-flow. Activities represent the basic functional elements of a behaviour, whereas their composition represents the logical order in which they can be executed. 
Work-flow composition is a key aspect of the service-oriented programming paradigm, which comes from the most credited business process language for Web Services, WS-BPEL.
In the following we present the definition of behaviour:

\begin{quote}
\emph{The behaviour of a service is the description of the service activities composed in a work-flow.}
\end{quote}
We distinguish three basic kinds of service activities:
\begin{itemize}
\item \emph{communication activities}: they deal with message exchange between services;
\item \emph{functional activities}: they deal with data manipulation;
\item \emph{fault activities}: they deal with faults and error recovery.
\end{itemize}

\subsubsection{Communication activities.}
Communication activities are called \emph{operations}. We inherit them from the Web Services Description Language specifications (WSDL)~\cite{wsdl-spec}. Operations can be divided into input operations and output operations. The former operations provide a means for receiving messages from an external service whereas the latter ones are used for message sending.
\begin{itemize}
\item Input operations:
\begin{itemize}
\item \textbf{One-Way}: it is devoted to receive a request message.
\item \textbf{Request-Response} : it is devoted to receive a request message and to send a response message back to the invoker.
\end{itemize}
\item Output operations
\begin{itemize}
\item \textbf{Notification}: it is devoted to send a request message.
\item \textbf{Solicit-Response}: it is devoted to send a request message and to receive a response message from the invoked service.
\end{itemize}
\end{itemize}
Such a categorization is also presented in~\cite{dumasInteraction,barros}, even if other authors consider only the single message exchange pattern (represented by One-Way and Notification) as in~\cite{ws-calculus}. Here we consider the Request-Response and Solicit-Response key interaction patterns for the service-oriented programming paradigm because they introduce specific issues from an architectural point of view. We will clarify such an aspect in Section \ref{sec:discussion}.

Output operations require the specification of a target endpoint to which the message has to be sent. At the level of behaviour such an endpoint abstractly refers to a service. Here, we call it \emph{receiving service}.

\subsubsection{Functional activities.} They allow for the manipulation of internal data by providing all the basic operators for expressing computable functions.

\subsubsection{Fault activities.} They allow for the management of faults. This is a fundamental aspect of service-oriented programming. The following list of basic activities for managing faults has been extracted from the experience we have developed on dynamic handling in SOCK and JOLIE~\cite{acsd08,ecows08} and other models and languages such as StAC~\cite{stac04}, SAGAS~\cite{sagas} and WS-BPEL.
\begin{itemize}
\item \textbf{Fault raising}: it deals with the signaling of a fault
\item \textbf{Fault handler}: it defines the activities to be performed when a fault must be handled
\item \textbf{Termination handler}: it defines the activities to be performed when an executing activity must be terminated before its ending.
\item \textbf{Compensation handler}: it defines the activities to be performed for recovering a successfully finished activity.
\end{itemize}

\subsection{Engine}
Here we present the concept of \emph{engine} that we introduced in SOCK for dealing with those aspects of the service-oriented programming paradigm related to the actual execution of a service into a network. To the best of our knowledge there is no clear and precise definition of engine as that which we presented in SOCK. The motivations of such a lack probably reside in the fact that it is usually considered an implementation detail that does not add anything relevant to service-oriented models. As far as service-oriented computing is concerned, engines are generally associated to those which execute WS-BPEL such as, for example, ActiveVOS~\cite{activeVOS} or the Oracle one~\cite{oracleBPEL}, but what about simple Web Services developed in Java, Python or .NET? Can we consider usual web servers such as IIS~\cite{iis}, Apache~\cite{apache} or Zope~\cite{zope} as service engines? Can we identify some characteristics on these applications and consider them basic features of the service-oriented programming paradigm? In our perception the answer is yes and this section is devoted to highlight the main features a service engine is characterized by. In general we say that an engine is a machinery able to create, execute and manage service sessions. A more detailed definition of engine will be provided at the end of this section, but we need to introduce some concepts first. Let us see the concept of session.
\subsubsection{Session.}
The definition of session follows:
\begin{quote}
\emph{A service session is an executing instance of a service behaviour equipped with its own local state.}
\end{quote}
A key element of the service-oriented programming paradigm is session identification. In general a session is identified by its own local state or a part of it. The part of the local state which identifies a session can be programmed and it is called \emph{correlation set}. Correlation sets is a mechanism provided by WS-BPEL and it has been formalized in SOCK, COWS and in~\cite{corrsets04}. We chose to allow for the definition of correlation sets even in JOLIE. In order to explain such a mechanism, let us introduce a simple notation, where a session is represented by a couple of terms $(P,S)$ where $P$ represents a behaviour in a given formalism and $S$ represents the local state here modelled as a function from variables to names. $S: Var\rightarrow Values$ where $Var$ is the set of variables and $Values$ the set of values\footnote{For the sake of brevity, here we consider both states and messages as a flat mappings from variables to values. The introduction of structured and typed values does not alter the correlation set insights presented in this section.}. Now, let us consider two sessions with same behaviour but different local states $S_1$ and $S_2$:
$$s_1:=(P,S_1)\ \ s_2:=(P,S_2)$$
We say that $s_1$ is not distinguishable from $s_2$ if $S_1=S_2$\footnote{Let $\Sigma$ be the set of states, we define $= :(\Sigma\times \Sigma)$ where $S_1 = S_2$ if $S_1(x)=S_2(x)$ and $Dom(S_1)=Dom(S_2)$}. Now, let us consider both ${\mathcal S_1}$ and ${\mathcal S_2}$ as a composition of states defined on disjoint domains:
$${\mathcal S_1}=S_{11}\oplus S_{12}\ \ \ \ \ {\mathcal S_2}=S_{21}\oplus S_{22}$$
where the operator $\oplus$ represents a composition operator over states\footnote{$\oplus:\Sigma\times \Sigma\rightarrow \Sigma$ where $S_1\oplus S_2(x) = S_1(x)\ if x\in Dom(S_1), S_2(x)\ if\ x\in Dom(S_2)\wedge x\notin Dom(S_1), undefined\ otherwise$}. Let us consider $S_{11}$ and $S_{21}$ as the correlation sets for $s_1$ and $s_2$ respectively. We say that $s_1$ is not distinguishable by correlation from $s_2$ if $S_{11}=S_{21}$.

Such a session identification mechanism differs from that of the object-oriented paradigm for objects. Indeed, in the object-oriented approach an object is always identified by the reference issued at the moment of the object creation. Such a reference cannot be used in service-oriented computing, because of its loosely coupled nature. This fact represents a major difference between the two approaches.

\subsubsection{Session management.}
Session management involves all the actions performed by a service engine in order to create and handle sessions. In order to achieve this task, a service engine provides the following functionalities:
\begin{itemize}
\item \textbf{Session creation}. Sessions can be created in two different ways:
\begin{itemize}
\item when an external message is received on a particular operation of the behaviour. Some operations can be marked as session initiators. When a message is received on a session initiator operation, a session can be started.
\item when a user manually starts it. A user can launch a service engine which immediately executes a session without waiting for an external message. We denote such a session as \emph{firing session}.
\end{itemize}
\item \textbf{State support}. The service engine also provides the support for accessing data which are not resident into session local states: the \emph{global state} and the \emph{storage state}. Summarizing, it is possible to distinguish three different kind of data resources, which we call \emph{states}, that can be accessed and modified by a session:
\begin{itemize}
\item a \textbf{local state}, which is private and not visible to other sessions. This state is deleted when the session finishes.
\item a \textbf{global state} which is shared among all the running sessions. This state is deleted when the engine stops.
\item a \textbf{storage state} which is shared among all the running sessions and whose persistence is independent from the execution of a service engine (e.g. a database or a file).
\end{itemize}
\item \textbf{Message routing.} Since a session is identified by its correlation set, the engine must provide the mechanisms for routing the incoming messages to the right session. The session identification issue is raised every time a message is received. For the sake of generality, in the service-oriented paradigm we cannot assume that some underlying application protocol such as WS-Addressing~\cite{ws-Addr} or other transport protocol identification mechanisms such as HTTP cookies~\cite{cookie} are always used for identifying sessions. We consider correlation sets as the representative mechanism for routing incoming messages. Its functioning can be summarized as it follows. A message can be seen as a function from variables to values: $M\in\Sigma$. As we have done for the states we can define a correlation set also for a message. Let us consider $M=M_1\oplus M_2$ where $M_1$ is the correlation set for the message $M$. We define the correlation function $c:Var\rightarrow Var$ which allows us to map message variables to state variables. We say that a message $M$ must be routed to the session $s$ whose state is $S = S_1\oplus S_2$ where $S_1$ is the correlation set, if:
$$\forall x\in Dom(M_1), c(x)\in Dom(S_1) \Rightarrow S(c(x)) = M(x) \vee S(c(x))\ is\ undefined$$ 
Informally, a message can be routed to a session only if its correlated data correspond to those of the session. The correlation function $c$ is the concrete means used by the programmers for defining correlation. For each incoming message it will be possible to define a specific function $c$ and the correlation set, which identifies the session, is indirectly defined by the union of the codomains of all the defined $c$ functions. It is worth noting that, if the correlation set is not correctly programmed, more than one running session could be correlated to an incoming message. In this case the session which has to receive the message  is non-deterministically selected.

\item \textbf{Session execution}. Session execution deals with the actual running of a created session behaviour equipped with all the required state supports. Sessions can be executed sequentially or concurrently. The majority of existing technologies share the idea that sessions are to be executed concurrently, but the sequential case allows for the controlling of some specific hardware resource which needs to be accessed sequentially. As an example, consider a cash withdrawal machine which starts sessions sequentially due to its hardware nature. 
We consider such an aspect fundamental from an architectural point of view because it can raise system deadlock issues if not considered properly, as we have shown in~\cite{sock-tech} by means of SOCK.
\end{itemize}

\subsubsection{Engine definition.}
Now we can provide our definition for service engine:
\begin{quote}
\emph{An engine is a machinery able to manage service sessions by providing session creation, state support, message routing and session execution capabilities.}
\end{quote}

\subsection{Service Description}
A service description provides all the necessary information for interacting with a service. Service descriptions are composed by two parts: interface and deployment.

\subsubsection{Interface.}
Service interfaces contain abstract information for performing compatibility checks between services, abstracting from low-level details such as communication data protocols and transports. We identify three different levels of service interface:
\begin{itemize}
\item \textbf{Functional.} It reports all the input operations used by the behaviour for receiving messages from other services or applications. An operation description is characterized by a name, and its request and response message types. If we trace a comparison with the Web Services technology, such an interface level is well represented by the WSDL specifications v1.1~\cite{wsdl-spec}.
At this level, only message type checks on the interface are required in order to interact with the service.
\item \textbf{Work-flow.} It describes the work-flow of the behaviour. In a work-flow, input operations could not be always available to be invoked but they could be enabled by other message exchanges by implementing a sort of high level application protocol. Thus, it is fundamental to know how a service work-flow behaves in order to interact with it correctly. In the Web Services technology such an interface could be provided by means of an Abstract-BPEL~\cite{wsbpel-spec} description; WSDL 2.0 specifications~\cite{wsdl-due} provide Message Exchange Patterns (MEP) which allows for the description of custom service interaction patterns, too. At this level, other approaches such as \emph{choreography} must be considered. Choreography languages, such as WS-CDL~\cite{wscdl-spec}, allow for the designing of a service system from a global point of view. In this case, a choreography can be exploited for describing the work-flow behaviour of a service by highlighting its role into the choreography. This is a wide area of investigation that involves works on contracts~\cite{gigio-contratti} and conformance~\cite{coord06}, but it is out of the focus of this paper. The reader interested in this topic may consult~\cite{conf/esop/honda,dmc05,conf/epew/BaldoniBMPS05,sefm08}.
\item \textbf{Semantics.} It offers semantic information about the service and the specific functionalities provided by it. It is usually provided by using some kind of ontology such as OWL-S~\cite{owl-s}.
\end{itemize}

Service interfaces are strictly related to service discovery, which is a key element of the service-oriented programming paradigm. Discovery issues are strictly related to search and compatibility check algorithms over interface repositories, also called \emph{registries}. These algorithms differ depending on which interface type is considered.
It is out of the scope of this paper to discuss the different methodologies used for implementing service discovery. However, here we want to highlight the fact that, at the present, some information related to service engine do not appear on the current interface standard proposals, such as the sequential execution modality or the correlation set of a service. These information are relevant and should be specified into the interface because they could influence other services in the system, as we have hinted in the previous section.

\subsection{Deployment}
The deployment phase is in charge of binding the service interface with network locations and protocols. A service, for example, could receive messages by exploiting the HTTP protocol or the SOAP over HTTP one, but the choice is potentially unlimited because new protocols may be created. Such a task is achieved by means of port declarations. There are two kind of ports: \emph{input ports} and \emph{output ports}. The former allows for the declaration of all the input endpoints able to receive messages exhibited by the service engine, whereas the latter bind target location and transport protocol to the \emph{receiving services} of the behaviour. In other words the output port allows for the concrete connection with the services to invoke. In general, we define a port as it follows:
\begin{quote}
\emph{A port is an endpoint equipped with a network address and a communication protocol joined to an interface whose operations will become able to receive or send requests. Ports which enables operations to receive requests are called input ports, output ports otherwise.}
\end{quote}
A service engine needs to be joined with deployment information in order to receive and send messages.

\subsection{Service}
Now, we are able to provide our definition of service which follows:
\begin{quote}
\emph{A service is a deployed service engine whose sessions animate a given service behaviour.}
\end{quote}
Such a definition is not enough if we consider the concrete execution of a service in a complex system. In order to complete our model we introduce the concept of \emph{container}. The definition of container is fundamental in the analysis of the system composition mechanisms described in the next section.

\begin{quote}
\emph{A service container is an application able to execute one or more services.}
\end{quote}

\section{Service Composition}\label{sec:composition}
Now that we have a definition of service and service container we can discuss the meaning of \emph{service composition}. The term composition discussed in Section \ref{sec:behaviour} concerns only the inner activities of a behaviour, which can be composed in a work-flow. Here we extend the usage of the term \emph{composition} at the level of service system, by showing the different mechanisms we experimented for putting services together into a system. To the best of our knowledge, this is the first attempt to classify different kinds of composition at the level of service system. These mechanisms became evident when we tried to implement the concepts of behaviour and engine, discussed in the previous section, in JOLIE, where we faced the challenge to offer to the programmer an easy way for building modular applications based solely on the concept of service. The following kinds of service composition emerged from our experience:
\begin{itemize}
\item simple composition
\item embedding
\item redirecting
\item aggregation
\end{itemize}
\begin{figure}
\centering
\includegraphics[scale=0.5]{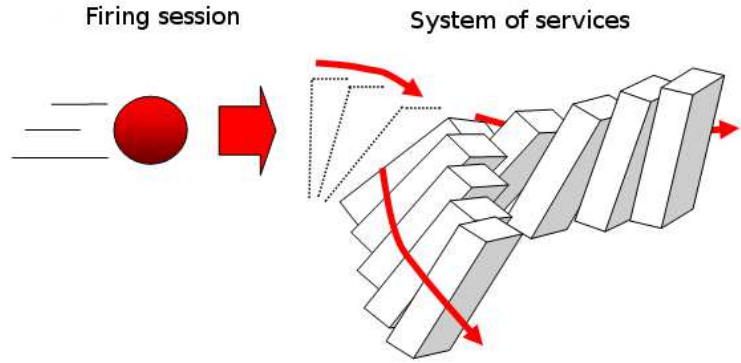}
\caption{A firing session starting a system of services.}\label{fig:domino}
\end{figure}
These composition techniques can be freely mixed together in order to obtain different architectures. The different composition techniques do not alter the functionalities of a service system but they allow for the engineering of the system architecture depending on the design needs. It is important to notice that, regardless of the kinds of composition that are used, one must ensure that at least one service present in the composition starts with a \emph{firing session}. This is necessary because the services not containing a firing session always have to wait for an input from an external process in order to be started. In order to clarify the concept of firing session let us consider Fig. \ref{fig:domino} where we represent the firing session as a ball which starts a set of domino cards. The spatial placement of the cards represents the dependencies among the services of the system started by the firing session. In the following we use the term \emph{client} for referring to a service that calls another service.

\subsection{Simple Composition}
\begin{quote}
\emph{Simple composition is the parallel execution of service containers into a network where they are able to communicate with each other.}
\end{quote}
In a simple composition the resulting system will behave accordingly to the behaviour of each service involved in. This is the most obvious way for composing services.

\subsection{Embedding}
\begin{quote}
\emph{Embedding is the composition of more than one service into the same container.}
\end{quote}
\begin{figure}
\centering
\includegraphics[scale=0.3]{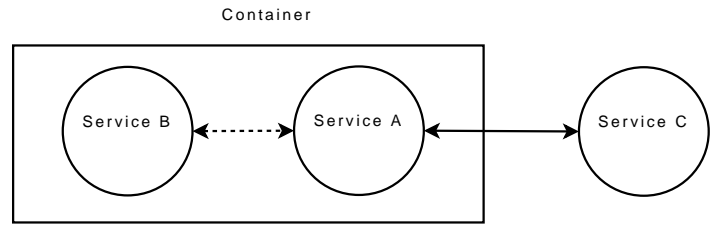}
\caption{Services A and B are embedded and their interactions do not exploit the network but they are performed within the container. Service C is not embedded and its interaction with service A exploits the network.}\label{fig:embedding}
\end{figure}
The main advantage of such a composition is the fact that embedded services could communicate with each other without using the network but exploiting inner communication mechanisms depending on the implementing technology (e.g.~RAM communications). 
Embedding is particularly suitable for increasing the level of granularity of a service-oriented system. Sending and receiving messages over a network indeed can be a strong limitation when we consider auxiliary services which provide basic functionalities (e.g.~services which provide mathematical functions, services that manage time and dates, etc.). By composing with embedding it is possible to use auxiliary services without deploying them into the network.
In Fig. \ref{fig:embedding} is represented the case where a service $A$ requires the functionalities of one other service $B$ but it does not make sense to deploy $B$ on the network and consequentially expose it to external processes. $A$ and $B$ are embedded and their interactions are performed within the container. On the other hand, $A$ can also interact via network with another service $C$ which is not embedded.

\subsection{Redirecting}
\begin{quote}
\emph{Redirecting allows for the creation of a \emph{master service} acting as a single communication endpoint to multiple services which are called resources.}
\end{quote}
\begin{figure}
\centering
\includegraphics[scale=0.3]{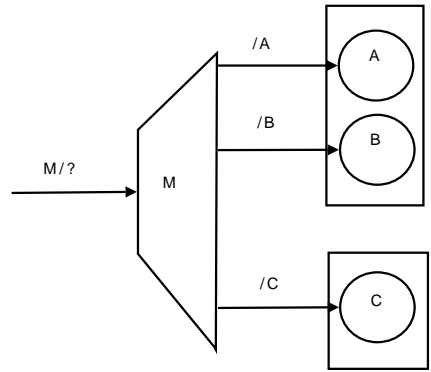}
\caption{Service $M$ redirects messages to services $A$, $B$ and $C$ depending on the target destination of the message ($M/A$, $M/B$ or $M/C$).}\label{fig:redirecting}
\end{figure}
In redirecting a master service receives all the messages of a system and then forward them to system services.
It is obtained by binding an input port of the master service to multiple output ports, each one identifying a service by means of a \emph{resource name}. The client will send messages to the master service specifying the final resource service to invoke. 
The main advantages of such an approach are:
\begin{itemize}
\item the possibility to provide a unique access point to the system clients. In this way the services of the system could be relocated and/or replaced transparently w.r.t. the clients;
\item the possibility to provide transparent communication protocol transformations between the client and the master and the master and the rest of the system.
\end{itemize}
In order to understand the second advantage better, consider Fig. \ref{fig:redirecting} and suppose that $A$ speaks a certain protocol $p_{a}$. Now suppose that a client needs to interact with $A$, but it does know only a different protocol: $p_{m}$. The client could then call $M$ with destination $M/A$ using protocol $p_{m}$ (known by $M$), and leave to $M$ the task of transforming the call message into an instance of $p_{a}$ before sending it to $A$.

\subsection{Aggregation}
\begin{quote}
\emph{Aggregation is a redirecting composition of services whose interfaces are joined together and published as unique.}
\end{quote}
Aggregation deals with the grouping of more services under the same interface. It is similar to redirecting, but the resource services are not visible from the point of view of the client. The client sees a unique service, the master one, which exhibits an interface by providing the functionalities of the resource services.
Differently from redirecting, which maintains the different interfaces of each composed service separated, in this case the client looses the details of each single service used behind aggregation.
The main advantage of such a composition approach deals with the possibility to completely hide the system components to the client.

\begin{figure}
\centering
\includegraphics[scale=0.3]{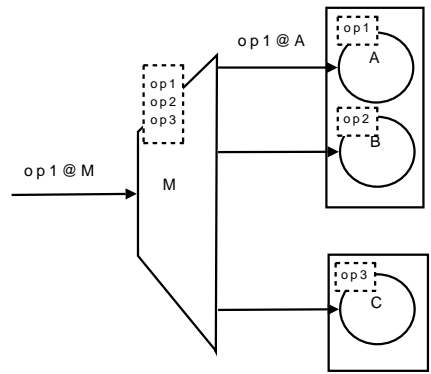}
\caption{In aggregation the master publishes the union of all the service interfaces it aggregates. Interfaces are here represented with dotted rectangles. The message on operation $op1$ to service $M$ is actually redirected to service $A$.}\label{fig:aggregation}
\end{figure}

\subsection{Dynamic System Composition}
The aforementioned service composition techniques (simple, embedding, redirection and aggregation) can be used both statically and at runtime. In the static case all the services are composed before their execution and the composition never changes during the execution of all the system. On the contrary, if the composition of the system changes at runtime, we say that the system is composed \emph{dynamically}.
Dynamic composition is strictly related to the concept of \emph{service mobility}. Service mobility deals with the representation of a service in some data format, its transmission from one service to another and then its execution in the service container of the receiver. It is worth noting that here we do not consider the case of running service migration but only the case of static service behaviour mobility.
In the following we briefly describe some case of dynamic composition:

\subsubsection{Dynamic embedding.} Let us consider a service which needs to receive software updates for a certain functionality. One may encapsulate that functionality in an embedded service. Then, when a software update is issued, the embedder service may unload the embedded one, receive the updated service to embed and dynamically embed the received service.

\subsubsection{Dynamic redirecting and aggregation.} Let us consider the case that a resource service faults or needs maintenance without affecting the service availability from the client point of view. It is sufficient to install a spare part resource service and registering it to the master service in place of the faulty one.

\section{Discussion}\label{sec:discussion}
The definitions of service and service composition shown so far are the key elements of the service-oriented programming paradigm. Such a paradigm becomes evident when building a system of services where it is possible to construct distributed system architectures by following new design patterns. In this section we present and discuss some of them.

\subsubsection*{Request-Response.}
The Request-Response interaction pattern plays a fundamental role when designing a service system architecture. Such a message exchange between two peers can be performed in two different ways: by means of a \emph{callback} or a Request-Response. In the former case both services exhibit a port for receiving the message, whereas in the latter case only the port of the Request-Response receiver is exhibited and the response message is sent on the same channel\footnote{It is out of the scope of this paper to provide a precise definition of channel. For the sake of our discussion, it is sufficient to informally consider a channel as a socket or an abstract channel like the one described in the WS-Addressing specification.} on which the request message is received.
\begin{figure}
\centering
\includegraphics[scale=0.3]{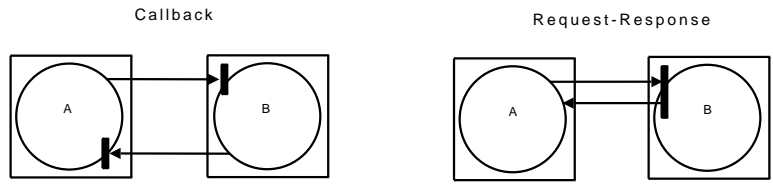}
\caption{Callback and Request-Response configurations. Black rectangles represent ports.}\label{fig:requestresponse}
\end{figure}
In Fig. \ref{fig:requestresponse} we graphically show the differences between a callback configuration and a Request-Response one. The main difference is that in the callback configuration the service $A$ must be aware of the fact that it has to receive the response message in a specific port. In other words, service $A$ must be aware of the work-flow behaviour of $B$, so it needs to know the work-flow interface of $B$ in order to interact with it. In the Request-Response case, it is sufficient for $A$ to know the functional interface of $B$ where the RequestResponse operation is declared in order to interact with it. Such architectural difference motivated us to consider both the approaches as fundamental for the service oriented programming paradigm. Moreover, the Request-Response pattern raises interesting issues from the fault handling point of view but, for the sake of brevity, we do not discuss them here and we invite the interested reader to consult~\cite{acsd08,ecows08}.

\subsubsection*{Web client-server pattern.}
It is easy to model the web client-server pattern in the service-oriented programming paradigm. A web server is a service deployed on an HTTP port, where it exhibits some general purpose Request-Response operations, such as \emph{GET} and \emph{POST}. Web servers usually manage session identification by means of cookies or query strings, which can be easily modelled with the correlation set mechanism.

\subsubsection*{Slave service mobility and Master service mobility.}
Here, we show two design patterns we experimented: slave service mobility and the master service mobility. We consider \emph{slave} the service which provides simple functionalities by means of Request-Response operations and \emph{master} the service which makes use of multiple slave services in its work-flow behaviour. In the \emph{slave service mobility} pattern the slave services are moved and embedded to the master service container, whereas in the \emph{master service mobility} the master service is moved and embedded into the container of the slave.
\begin{figure}
\centering
\includegraphics[scale=0.3]{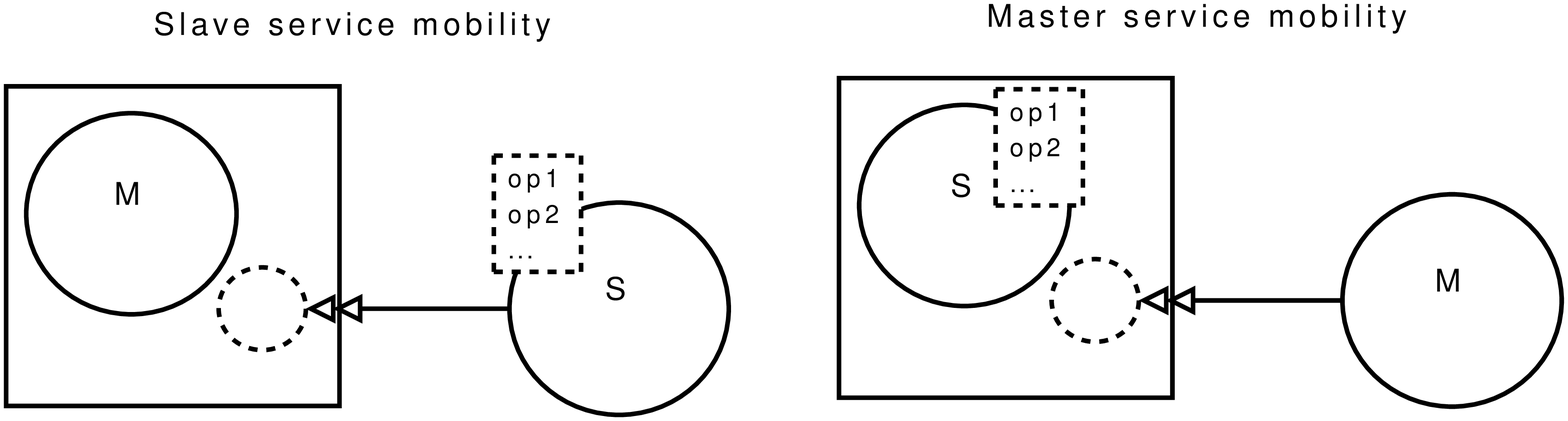}
\caption{Slave service mobility and Master service mobility.}\label{fig:smmobility}
\end{figure}
In the following we explain both the patterns by referring to Fig. \ref{fig:smmobility}
\begin{itemize}
\item \emph{Slave service mobility.} Consider the case in which a service $M$ (the \emph{master service}) defines a work-flow that is dependent on some functionalities that cannot be provided statically before execution time. $M$, instead, needs to obtain these functionalities at runtime and to use them. In order for this to work, $M$ must define an appropriate output port  for the functionalities it is looking for. Then, $M$ asks a service repository for downloading the functionalities it needs. The repository sends a service $S$ to $M$, and $M$ dynamically embeds $S$. $M$ has now access to the functionalities offered by $S$, the \emph{slave service}, and exploit them to complete its work-flow. 
\item \emph{Master service mobility.} 
Consider the case in which a service $S$ (the \emph{slave service}) possesses the functionalities that are needed for the actualization of a work-flow, but the work-flow cannot be provided statically before execution time. $S$ needs to obtain the work-flow at runtime and to execute it, ensuring that the work-flow makes use of the functionalities provided by $S$. We exploit such a pattern for implementing the SENSORIA automotive scenario~\cite{Sensoria} where a car experiments a failure and starts a recovery work-flow for booking some services such as the garage, the car rental and the truck one. We implemented it with a slave service on the car and the master work-flow which is downloaded from the car factory assistance service. In this way we obtain that the recovery work-flow can be changed and maintained by the assistance car service without updating all the car softwares periodically and, at the same time, we guarantee transaction security by isolating some functionalities such as the bank payment, into the slave service of the car. In this case the downloaded work-flow is able to search for all the services it needs but it relies upon the slave service car functionalities for the payment.
\end{itemize}


\subsubsection*{SoS: Service of services pattern.}
The SoS pattern exploits both dynamic embedding and dynamic redirecting. A service is embedded at run-time if a client performs a resource request to the master service. In this case the master service embeds the requested service by downloading it from a repository and make it available to the client with a private resource name. From now to the end, the client will be able to access to its own resource by simply addressing its requests to the resource name it has received. The main advantage of this approach is that we can provide an entire service as a resource to a specific client instead of a single session of a service. We exploit this pattern for implementing the SENSORIA finance case study~\cite{Sensoria} which models a finance institute where several employees works on the same data. We use the SoS pattern for loading a service for each employee which maintain its private data and, at the same time, can offer a set of functionalities. The main advantage in this approach is that each functionality offered to the employee is able to open a session on its own service thus obtaining a private complex resource made of a set of service sessions.

\section{Conclusions}\label{sec:conclusions}
We have presented the results of our experience in investigating service-oriented applications both from a theoretical and a practical point of view. It is our opinion that a service-oriented programming paradigm exists and that it is characterized by the concepts of behaviour, session, session execution, correlation set, engine, interface, deployment, service, container, embedding, redirecting and aggregation. Upon these key concepts we experimented some interesting design patterns from an architectural viewpoint, such as the slave service mobility, the master service mobility and the SoS. They probably are only some of the possible design patterns which can be built with the service-oriented programming paradigm. In the future we will continue to investigate the service-oriented programming paradigm by also considering choreography languages as complementary means for designing service-oriented systems. As far as JOLIE is concerned, we will continue in its development by studying and implementing all the necessary mechanisms for obtaining the aggregation service composition pattern and tools for improving usability and property checking.

\section*{Acknowledgments}

Research partially funded by EU Integrated Project Sensoria, contract n.~016004.

\bibliographystyle{eptcs}
\bibliography{biblio}

\end{document}